\documentclass[runningheads]{llncs}
\usepackage{graphicx}
\usepackage{comment}
\usepackage{amsmath}
\usepackage{color,soul}

\def\v#1{\mathbf{#1}}
\def\m#1{\mathnormal{#1}}

\begin{document}
\title{Surgical Video Motion Magnification with Suppression of Instrument Artefacts}
%
%
\author{Mirek Janatka\inst{1}
	\and Hani J. Marcus\inst{1,2} 
	\and Neil L. Dorward 
\inst{2} \and
Danail Stoyanov \inst{1}} 

\authorrunning{M. Janatka et al.}
%
\institute{Wellcome/EPSRC Centre for Interventional and Surgical Sciences,
	University College London, London, UK \and
Department of Neurosurgery, The National Hospital for Neurology
and Neurosurgery, Queen Square, London, UK\\
\email{mirek.janatka@ucl.ac.uk}}
\maketitle              
\begin{abstract}
Video motion magnification can make blood vessels in surgical video more apparent by exaggerating their pulsatile motion and could prevent inadvertent damage and bleeding due to their increased prominence. It could also indicate the success of restricting blood supply to an organ when using a vessel clamp. However, the direct application to surgical video could result in aberration artefacts caused by its sensitivity to residual motion from the surgical instruments and would impede its practical usage in the operating theatre. By storing the previously obtained jerk filter response of each spatial component of each image frame - both prior to surgical instrument introduction and adhering to a Eulerian frame of reference - it is possible to prevent such aberrations from occurring. The comparison of the current readings to the prior readings of a single cardiac cycle at the corresponding cycle point, are used to determine if motion magnification should be active for each spatial component of the surgical video at that given point in time. In this paper, we demonstrate this technique and incorporate a scaling variable to loosen the effect which accounts for variabilities and misalignments in the temporal domain. We present promising results on endoscopic transnasal transsphenoidal pituitary surgery with a quantitative comparison to recent methods using Structural Similarity (SSIM), as well as qualitative analysis by comparing spatio-temporal cross sections of the videos and individual frames.
\end{abstract}
\section{Introduction}
In endoscopic surgery visualising blood vessels is a common challenge as they are often beneath the tissue surface or indistinctive from the surface texture. Major complications can result from instruments causing inadvertent damage and hence bleeding \cite{feldman1961problem} due to vessel imperceptibility that leads to surgical error. In severe cases, such bleeding can place the patient at risk of death if it cannot be controlled or potentially lead to post-operative problems, that would require additional surgery to address \cite{marietta2006pathophysiology}. The major challenge and cause for such problems in endoscopic procedures is that subsurface vessels cannot be visualised directly or detected through touch and palpation \cite{stoyanov2012surgical}. Various schemes for detecting and avoiding subsurface vasculature have been explored including Augmented Reality (AR) that fuses preoperative patient imaging with the surgical video, as well as novel optical imaging \cite{clancy2020surgical} and intraoperative ultrasound \cite{cleary2010image}. Similar schemes have also been used to determine if blood flow continues to pulse into a region, after being obstructed by a clamping mechanism \cite{norat2019application}.
\begin{figure}
	\centering
	\includegraphics[width=\textwidth]{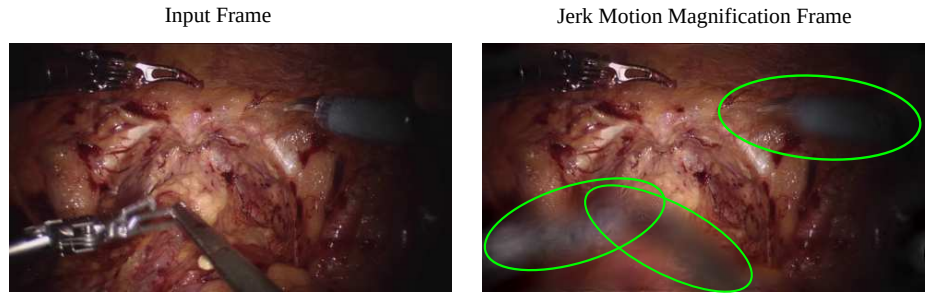}
	\caption{Demonstrating aberrations from surgical instrument motions using video motion magnification via frame comparison. Left) Original video frame. Right) Aberrations generated jerk-based motion magnification, outlined by the green elliptical annotation \cite{janatka2018higher}}	
	\label{fig:BlurExample}
\end{figure}
However, all have some drawbacks in terms of surface registration accuracy, workflow additions, ergonomic problems and signal sensitivity \cite{bernays2010intraoperative}. Hence practically, the problem persists and can be a significant hurdle to the successful completion of many procedures
\cite{shander2007financial}.
Video motion magnification (VMM) \cite{Wadhwa2012,EVMsource,le2019seeing} has previously been proposed as a mechanism to aid vessel localisation in endoscopic video, without the requirement of additional \textit{in situ} hardware, surface registration or contrast agents \cite{mcleod2014motion,amir2015automatic,janatka2018higher}. VMM uses the existing variations in the endoscopic video stream that are minute and out of the range of the surgeon's perception and creates a synthetic video where such motions are perceivable. The generated video characteristics can be temporally selected, so that reoccurring motion within a certain temporal frequency range can be processed exclusively, which allows for 
motions of known occurrence to be selected \cite{Wadhwa2012,EVMsource}. In the case of vessel localisation, it is possible to isolate and amplify motion due to the heart rate (readily available in the OR), that governs the periodicity of vessel distension from the pressure wave that is generated from the heart. VMM assumes spatial consistency, effectively treating every point on the image as a time series, and requires a static view for initialisation. This creates limitations and while VMM has been demonstrated in endoscopic third ventriculostomy and robotic prostatectomy videos, it has not been effectively clinically translated or adopted \cite{mcleod2014motion}. VMM has also been suggested for other clinical applications, such as with video otoscopy \cite{janatka2017examining}.

One challenge for VMM in surgery is that other motions within the scene can cause aberrational distortions and can be disorientating to surgical view. An alternative method of using VMM was proposed in robotic partial nephrectomy, where respiratory motion is present and needs be accounted for. Rather than use VMM directly a colour map representation was generated from the raw VMM video of where pulsation was used to located vessels as an aid to assist in registration of preoperative patient data \cite{amir2015automatic}. Attempts to deal with large motion presence in motion magnified videos have been suggested by segmentation which is not practical in an operation \cite{elgharib2015video,kooij2016depth}. Yet, recent developments in temporal filtering have allowed for different components of motion, such as acceleration and jerk, to be selected based upon a principal oscillation frequency. This allows for VMM to leverage a higher band pass of frequencies than the band under investigation to exaggerate motion within video \cite{janatka2018higher,zhang2017video,takeda2018jerk,takeda2019video}. In our previous work, we used the third order of motion (jerk) was utilised to exclude motions from respiration and transmitted motion from larger arteries, whilst motion magnifying blood vessels in surgical video, with a filter designed around the pulse rate. This approach reduced blur distortion caused by the large motions within the scene, whilst still permitting motion magnification of vessel distension, as the pulsatile waveform contains jerk characteristic \cite{janatka2018higher}. However, this jerk filter is not able to prevent the generation of motion blur aberrations from the presence of instruments moving within the scene, making the synthesised video unsuitable for surgical guidance (as shown in Fig.\ref{fig:BlurExample}). For VMM to be a viable option for surgical intervention it must be usable with instrument motion presence or its function would be limited to just observational usage.

In this paper, we propose a technique that would allow for motion magnification to be left unaffected by the introduction of tool motion to the surgical camera's field of view. To operate, it simply requires the known heart rate and a brief sample of the view that is uninterrupted by instrument motion. It maintains the spatial consistency assumption of a fixed view point. We demonstrate this method on four cases of endoscopic video of transnasal transsphenoidal surgery, providing both qualitative and quantitative comparison to a prior method.

\section{Methods}
\subsection{Motion Magnification}
VMM operates by spatially decomposing video frames into local frequency components using Complex Steerable Pyramid (CSP) \cite{portilla2000parametric,freeman1989steerable} which uses different Gabor-like wavelets ($\psi$) to decompose images at varying scales and orientations representations notated by $\m S$. 

\begin{equation}
\tilde{I}(\v x,t) = \sum {\m S}(\v x,t)* \psi + \epsilon(\v x,t)
\end{equation}
Where $\tilde{I}(\v x,t)$ represents a CSP reconstructed video frame at time $t$, with intensity values at $\v x = (x,y)$ pixel value, $*$ notes a convolution operation. As well as the band-pass of $S$ a high-pass and low-pass residual ($\epsilon(\v x,t)$) that is unalterable is also required for the reconstruction.
As the information held in $\m S(\cdot)$ are complex conjugates representing local frequency information, local phase can be attained. As the local motion is related to the local phase ($\phi(\v x,t)$) via the Fourier shift theorem, motion analysis and modulation can be performed by filtering and manipulating the local phase of the video over time. As shown in our previous work \cite{janatka2018higher}, higher order of motion magnification utilises a temporally tuned third order Gaussian derivative to detect jerk motion $\m D(\v x, t)$ within a certain pass-band of temporal frequencies. It exploits the linear relationship of convolution to gather third order of motion from local phase using a third order Gaussian derivative:
\begin{equation}
\m D(\v x, t) := \frac{\partial^{3} G_{\sigma}(t)}{\partial t^{3}} * \phi(\v x, t) = \frac{\partial^{3} \phi(\v x, t)}{\partial t^{3}} * G_{\sigma}(t)
\end{equation}
Where $\sigma$ is the standard derivation of the Gaussian ($G$) derivative and is defined $\sigma = \frac{fr}{4\omega}$ \cite{lindeberg1990scale}, where $\omega$ is the temporal frequency of interest, in this case the heart rate from the electrocardiogram. $fr$ denotes the sampling rate of the video. The Gaussian derivative convolution is applied to the time series for all scales and orientations of each pixel ($\v X$) from the CSP representation.
By detecting local jerk motion in a video, it can be exaggerated by an amplification factor $\alpha$ to generate $\hat{\m S}(\cdot)$. 
\begin{equation}
\hat{\m S}(\v x,t)= A(\v x,t)e^{i(\phi(\v x, t) + \alpha{\m D}(\v x,t))}
\end{equation}
Where $A$ is the amplitude and $\phi$ is the phase of that particular local frequency at $\v x$ at time $t$ with respect to the $\m S$ band's orientation and scale. The summations of which from the various scales and orientations can reconstruct the motion magnified frame $\hat{I}(\v x,t)$. 
\subsection{Tool Motion Artefact Suppression Filter}
Assuming the endoscope is statically positioned and there are no surgical instruments within the scene, the response of the jerk filter applied to the video feed can be anticipated as:
\begin{equation}
\m D(\v x,{t}) = \m D(\v x,{t}\bmod{\frac{1}{\omega_{c}}})
\end{equation}
Where $\bmod$ is the modulo function and $\frac{1}{\omega_{c}}$ is the time period of the cardiac cycle, reported from the electrocardiogram. ${\omega_{c}}$ is also the $\omega$ value used to determine $G(t)$ from Eq. 2.
However, in reality, due to sampling quantization and subtle variations in the heart rate, perfectly aligned repetition rarely occurs. Therefore these values are used as a guide for creating the filter with an offset range, which can then be loosened by a scaling factor. Taking these stored tool free scene readings to be $\m TL(\v x)$:
\begin{equation}
\m TL (\v x) = {\m D}(\v x,0), {\m D}(\v x,1), \dots {\m D}(\v x,\frac{1}{\omega_{c}})
\end{equation}
To generate the offset range for the filter, the variability of $\m{TL}(\v x)$ can be found as $R(\v x)$: 
\begin{equation}
R(\v x) = \frac{ max(\m TL(\v x) ) - min(\m TL(\v x) )}{2} 
\end{equation}
This allows the comparator filter to be:
\begin{equation}
\m D(\v x,{t}\bmod{\frac{1}{\omega_{c}}}) \pm \beta R(\v x) 
\end{equation}  
Where $\beta$ is a scaling factor that can widen the filter further.
To operate within the VMM, the comparator has to act as a switch so that the motion magnification effect can be deactivated. Therefore we define the state of $\chi(\v x, t)$ as:
\begin{equation}
\begin{array}{ll}  1 = [{\m D}(\v x,t) <  {\m D}({\v x,t}\bmod{\frac{1}{\omega_{c}}}) + \beta R(\v x)]
\land  [{\m D}(\v x,t) >  {\m D}({\v x,t}\bmod{\frac{1}{\omega_{c}}}) - \beta R(\v x)] 
\\
0 = else                 
\end{array}
\label{eq:ToolThresLogic}
\end{equation}
Where $\land$ is a logic $and$ function. This consideration can be shown in the CSP magnified band $\m S$ as $\hat{\m S}(\cdot)^{*}$, with the threshold value $\beta$ being pre-assigned with $\chi_{\beta}$. 
 \begin{equation}
\hat{\m S}(\v x,t)^{*}= A(\v x,t)e^{i(\phi(\v x, t) + \alpha\chi_{\beta}(\v x, t){\m D}(\v x,t))}
\end{equation}
Which can be used to reconstruct $\hat{\m I}(\v x,t)^{*}$, a motion magnified video without blur distortion from tool motion.

\subsection{Synthetic Example}
\begin{figure}[!ht]
	\centering
	\includegraphics[width=\textwidth]{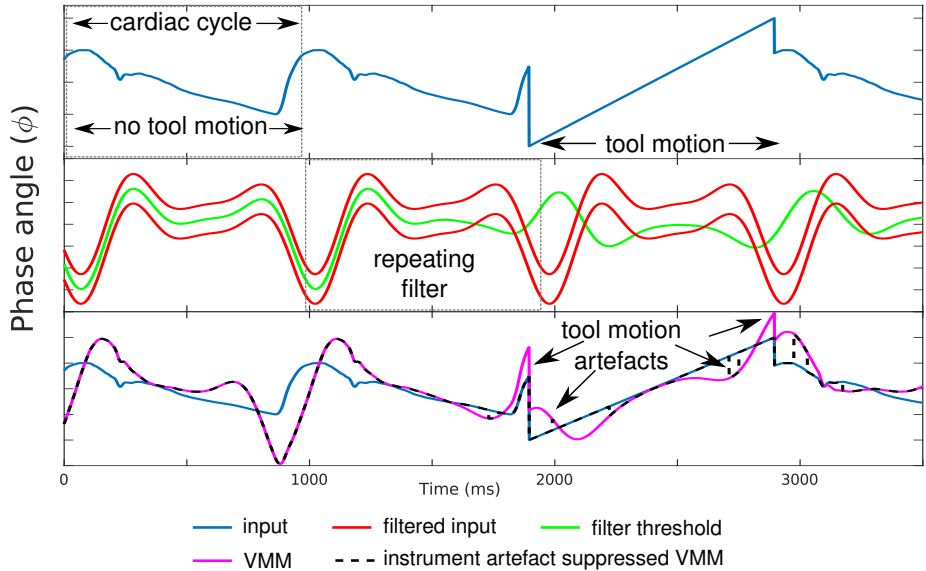}
	\caption {Synthetic one dimensional explanation of the instrument motion suppression filter. The top plot shows the observed cardiac motion with a tool passing over it. The middle plot shows the filtered component of the motion signal and the generation of the suppression filter. The bottom plot shows the difference the suppression has on the motion magnification output signal.}	
	\label{fig:SynthResults}
\end{figure}

To better visualise how the tool motion artefact suppression filter (TMASF) works, Fig. \ref{fig:SynthResults} shows a demonstration on a synthetic arterial displacement profile \cite{VirtualPulseWave}, shown as a phase reading that could be taken from a single pixel from an arbitrary $S(\cdot)$, where vessel motion exists. After just under two cardiac cycles, a tool passes across the point, denoted by ``tool motion" that alters the phase reading. The reading then returns to that of the vessel motion as before. 
The response of this displacement profile from the jerk filter \cite{janatka2018higher} is shown in the middle plot of Fig. \ref{fig:SynthResults} in green. After a single cardiac cycle, the TMASF can be constructed, the bounds of which are shown in red. So long as the jerk filter response stays within these bounds, $\chi(\cdot)$ is equal to 1 and the amplification is performed. 

However, if the jerk filter response moves outside these bounds, $\chi(\cdot)$ is equal to 0 for that particular pixel around that moment and the amplification effect is nullified. This can be seen around the time the tool motion is present. The bottom plot of Fig. \ref{fig:SynthResults} shows the resultant amplified signal, using both the TMASF filter (dashed black curve) and without (magenta curve). As shown, the amplification generates tool motion artefacts on prior method where tool motion is present. The TMASF reduces such artefacts, but is not immune to them. The extent to which they are created can be reduced by decreasing the $\beta$ value, which would essentially bring the red lines in Fig. \ref{fig:SynthResults} closer together. Yet, this could also be detrimental to the amplification effect, due to the imperfections in regularity and sampling. 
\section{Results}
\subsection{Experimental Setup}
To verify if the TMASF works in surgical videos, we performed a proof of concept study using retrospective data (IDEAL Stage 0) \cite{sedrakyan2016ideal}, applying the filter to a series of patient cases (n=4) that underwent endoscopic transnasal transsphenoidal surgery. The study was registered with the local institutional review board, patients provided their written consent, and videos were fully anonymised.
Each video was processed with the TMASF at three $\beta$ threshold scalings ($\beta$~= 1, 3 and 5) and without, using the previous jerk motion magnification without the TMASF \cite{janatka2018higher}, using 8 orientations and quarter octave CSP, from hereafter referred to as VMM.

 All samples have a brief few seconds before surgical instruments are visible in the scene for the filter to initialise. This would be a reasonable prerequisite requirement and perhaps automatically possible with a surgical robot as it is aware when the system is stationary. For all videos the magnification factor was set to $\alpha = 10$. The resolution of the capture videos were 1280 x 720, however were cropped to a 720 x 720 pixel square to account for the visible area of the video. The videos were scaled down by a half for quicker offline processing. Acquisition of the endoscope was at 25 fps. For quantitative analysis and comparison Structure Similarity Image Matrix index (SSIM) was used to compare the magnified frames to the corresponding input frames. The closer to 1 the similar the frame is to the input, meaning the less noise or motion magnification effect has been generated. To verify that the motion magnification effect is still operational, comparative spatio-temporal cross sections from places of interest are compared across all videos. Additionally, all videos are supplied in the supplementary material.
 \subsection{Results and Analysis}
\begin{figure}[!ht]
	\centering
	\includegraphics[width=\textwidth]{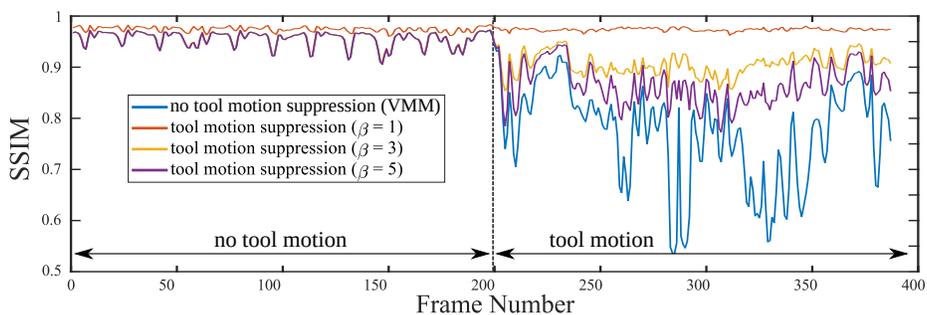}
	\caption {Frame-wise comparison of SSIM for all four motion magnified videos (Case 4)}	
	\label{fig:SSIMcaseFour}
\end{figure}

\begin{figure}[!ht]
	\centering
	\includegraphics[width=0.92\textwidth]{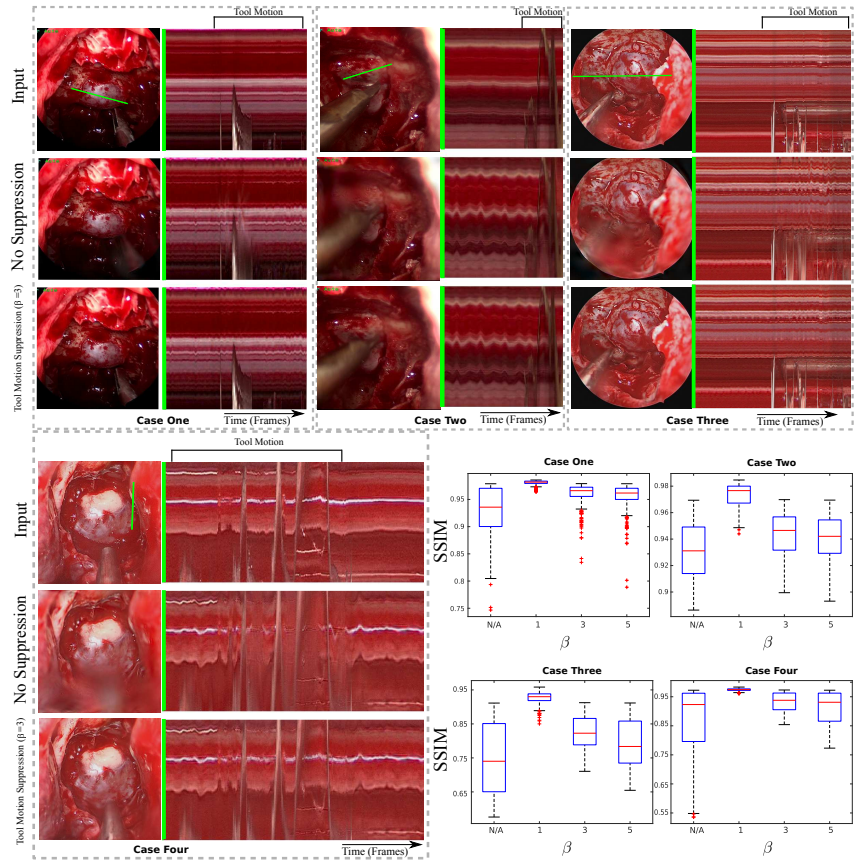}
	\caption {Qualitative (input, VMM and TMASF $\beta$ = 3 - green line on image frame indicates sample site for spatio-temporal cross section) and quantitative (using SSIM: VMM and TMASF $\beta$ = 1,3 and 5 ) comparisons of all four cases.}	
	\label{fig:VideoResults}
\end{figure}

As a comparative example of how the TMASF functions, the frame-wise SSIM result for each scaling factor of the TMASF and VMM can be seen in Fig. \ref{fig:SSIMcaseFour}, that depicts the result from case four, and is a common representation for all cases. 
The ``tool motion" region shows how the various $\beta$ values of the TMASF perform. The VMM curve shows the impact tool movement has on the generated video, with SSIM values dropping to as low as 0.55 and are volatile for this duration, with rarely reaching the lowest reading from the ``no tool motion" region. Similarly to the prior ``no tool motion" section, TMASF $\beta$ = 1 shows that the motion magnification is impinged by the suppression filter at this $\beta$ level. The other two $\beta$ values show a drop in SSIM readings, compared to the ``no tool motion" region, however are not as severe as the VMM curve, with $\beta$ = 5 reaching an offset of -0.05 compared to $\beta$ = 3 at most. These results suggest that the TMASF works, on $\beta$ levels 3 and higher. To report a fuller quantitative performance of the TMASF, box plots from the SSIM reading of the entire videos are shown on the right bottom corner of Fig.\ref{fig:VideoResults}. A running trend of the order of the medians from all cases for the different scaling values can be seen, with VMM (N/A) being lowest, followed by $\beta$ = 5, then $\beta$ = 3 and highest being $\beta$ = 1.
The images in Fig.\ref{fig:VideoResults} shows qualitative comparisons of each of the four cases with select frames from VMM and TMASF $\beta$ = 3 to the input video. Next to each frame is a spatio-temporal cross section, that is taken from each respective video (indicated by a green line) and shows how the pixels change in that location over time. By looking at the frames from VMM and TMASF $\beta$ = 3 it can be seen that blur distortion is reduced and that the TMASF makes the video clearer to view. The spatio-temporal cross sections show that physiological motion magnification is present in all cases, however for the TMASF structure can still be seen where tool motion is present (similar to the image in the input cross section), whilst is lost in the VMM cross section. This suggests that TMASF successfully reduces aberrations from VMM videos whilst retaining the desired motion magnification effect of exaggerating the motion of the physiology. In general, the results suggest that there is a trade-off between impinging the motion magnification effect and permitting aberration occurring from instrumentation motion, depending on the $\beta$ value used with TMASF.

\section{Discussion}

In this paper, we have proposed a filter constructed from local phase information collected prior to the insertion of surgical tools into the surgical field of view for motion magnification in endoscopic procedures. This approach can prevent aberration from being created due to instrument motion and hence allow more clinically usable surgical motion magnification augmentation, such as critical structure avoidance and assistance in vessel clamping. 

We have shown that the proposed filter can reduce amplification of motion due to tools on example videos from endoscopic neurosurgery where the camera and surgical site are confined and do not move too much. This is an important consideration because our filter would need re-initialising if the endoscope is moved or after large changes to the surgical scene. Yet this could be performed quickly, as the initialisation period is the length of a heart beat but more work is needed to detect and automate any re-initialisation strategy. 

The application of the TMASF is causal and is possible to combine with a real-time system as it is only reliant on past information. Further work is needed to consider user studies investigating how augmented visualisation of motion magnification can be presented to the clinical team to access the risk of cognitive overload and inattention blindness \cite{marcus2015comparative}. Both the motion magnification factor $\alpha$ and TMASF $\beta$ value are variables that can be tuned to create the optimal synthesized video, however, it is not essential that the raw video is used alone. TMASF could assist in existing visualisation approaches to only show physiological motion \cite{amir2015automatic}. Additionally, the ability to segregate physiological and non-physiological motion from a surgical scene could assist in surgical instrument tracking and anatomical segmentation tasks, as well as in other modalities of medical imaging that utilise motion estimation. 

\subsection{Acknowledgements}
The work was supported by the Wellcome/EPSRC Centre for Interventional and Surgical Sciences (WEISS) [203145Z/16/Z]; Engineering and Physical Sciences Research Council (EPSRC) [EP/P027938/1, EP/R004080/1, EP/P012841/1]; The Royal Academy of Engineering [CiET1819\slash2\slash36]; Horizon 2020 FET (GA 863146) and the National Brain Appeal Innovations Fund.

\end{document}